# Bio-optical characterization using Ocean Colour Monitor (OCM) on board EOS-06 in coastal region

Anurag Gupta*[1], Debojyoti Ganguly[1], Mini Raman[1], K. N. Babu[1], Syed Moosa Ali[1], Saurabh Tripathi[2]

[1]Biological Oceanography Division, Space Applications Centre (ISRO), Satellite road, Ahmedabad, Gujarat, India 380015.

[2]Met & Ocean Cal-Val Division, Space Applications Centre (ISRO), Satellite road, Ahmedabad, Gujarat, India 380015.

Corresponding author(s). *E-mail(s): anuraggupta@sac.isro.gov.in

# Bio-optical characterization using Ocean Colour Monitor (OCM) on board EOS-06 in coastal region

## Abstract

In ocean colour remote sensing, radiance at the sensor level can be modeled using molecular scattering and particle scattering based on existing mathematical models and gaseous absorption in the atmosphere. The modulation of light field by optical constituents within the seawater waters results in the spectral variation of water leaving radiances that can be related to phytoplankton pigment concentration, total suspended matter, vertical diffuse attenuation coefficients etc. Atmospheric correction works very well over open ocean using NIR channels of ocean colour sensors to retrieve geophysical products with reasonable accuracy while it fails over sediment laden and/or optically complex waters. To resolve this issue, a combination of SWIR channels or NIR-SWIR channels are configured in some ocean colour sensors such as Sentinel- OLCI, EOS-06 OCM etc. Ocean Colour Monitor (OCM)-3 on board EOS -06 was launched on Nov 26, 2022. It has 13 bands in VNIR (400-1010 nm range) with ~1500 km swath for ocean colour monitoring. Arabian Sea near Gujarat coast is chosen as our study site to showcase the geophysical products derived using OCM-3 onboard EOS-06.

# 1. Introduction

Remotely sensed ocean-colour observations using satellite sensors are widely utilized as an important component to understand the ocean. The ocean chlorophyll-a concentration, the primary constituent defining the ocean-colour program and the associated sensor configuration, is a measure of marine phytoplankton biomass. The Phytoplankton are responsible for approximately half of the global carbon uptake through photosynthetic process (Field *et al*., 1998). In response to the potential importance of phytoplankton in the global carbon cycle and the lack of comprehensive data, the international communities have established the satellite missions designed to acquire and produce high quality global ocean-colour data. One of the goals of launching a number of ocean-colour sensors aboard various satellites is to build a long-term, multi-sensor, multi-year, ocean-colour archive (IOCCG, 1999; McClain, 1998). The derived chlorophyll-a concentrations (in time and space) can be used to resolve inter-annual to decadal changes in oceanic phytoplankton biomass in response to global environmental changes.

Ocean colour remote sensing deals with quantitative estimation of the bio-optical properties from spaceborne measurements. The visible to near-infrared (VIS-NIR) region of the electromagnetic spectrum is of prime concern to study the ocean surface water column. Light interactions with optical constituents in the water column affect multiple scattering and absorption Consequently, the resultant signal from the water column reaches the satellite sensor in space. Thus, the remote sensing reflectance is derived by removing the absorption and scattering due to gaseous molecules and particles present in the atmosphere from top-of-atmosphere (TOA) radiance, followed by retrieval of bio-optical constituents using well-known bio-optical algorithms.

## *1.1 Indian Scenario*

India has a committed program on Ocean Colour. The first Ocean Colour Monitor (OCM), called Oceansat-1, was launched on May 26, 1999, on board the Indian Remote Sensing Satellite (IRS-P4) to collect valuable data set on the bio-optical parameters from space. Following the success of Oceansat-1 and to provide continuity of ocean-colour data, the Indian Space Research Organisation (ISRO) launched a second satellite, Oceansat-2, on September 24, 2009, with OCM-II to further enhance the study of oceans. Data from Oceansat-1 OCM and Oceansat-2 OCM, has been operationally utilized for the identification of Potential Fishing Zone (PFZ) and its forecast by the Indian National Centre for Ocean Information Services (INCOIS). Apart from the operational use, data from OCM1 and OCM2 has been utilized for various biological and geo-physical applications like studying coastal processes, atmospheric aerosol radiative forcing, physical-biological coupled processes, primary and secondary productivity estimates, sea surface nitrate mapping, species specific fishery forecast, etc. OCM1 and OCM2 comprised of eight spectral bands in the visible and near-infrared (VNIR) region of electromagnetic spectrum to collect data on the objects of interest (targets) as well as for atmospheric correction. OCM3, an improvement over OCM1 and OCM2, features indigenous Charge Coupled Device detectors with 13 channels. These cover the spectral range of 0.407 to 1.020 microns to meet the operational and scientific applications need. Table 1 provides a glimpse of the enhancements in OCM-3 over OCM 1 and 2.

*Table 1: Technical specifications of Ocean Colour Monitor onboard EOS-06*

| Band No. | OCM-1 (centre wavelength) | OCM-2 (centre wavelength) | OCM-3 (centre wavelength) |
|---|---|---|---|
| 1 | 412 nm | 412 nm | 412 nm |
| 2 | 443 nm | 443 nm | 443 nm |
| 3 | 490 nm | 490 nm | 490 nm |
| 4 | 510 nm | 510 nm | 510 nm |
| 5 | 555 nm | 555 nm | 555 nm |
| 6 | 640 nm | 620 nm | 566 nm |
| 7 | *765 nm | *740 nm | 620 nm |
| 8 | *865 nm | *865 nm | 670 nm |
| 9 | | | 681 nm |
| 10 | | | 710 nm |
| 11 | | | *780 nm |
| 12 | | | *870 nm |
| 13 | | | *1010 nm |
| No. of bands | 8 bands | 8 bands | 13 bands |
| SNR (in visible bands) | ~300 | ~300 | 800-1000 |
| Bandwidth | 20 nm (Application bands)  *40 nm (Atmospheric correction bands) | 20 nm (Application bands)  *40 nm (Atmospheric correction bands) | 10/8 nm (Application bands)  *20 /40nm (Atmospheric correction bands) |

Globally, conventional, standard atmospheric correction procedure to were adopted to retrieve ocean remote sensing reflectances from top-of-the–atmosphere radiances in OCM-1 and OCM-2 data. These procedures used dark pixel approximation using NIR bands and single scattering approximation to determine aerosol radiance at NIR bands of OCM-1 and OCM-2. Bio-optical algorithms OC4 involving 4 bands 443, 490, 510 and 555 nm with globally tuned coefficients were adopted for the retrieval of chlorophyll-a concentration. Regional datasets were used to empirically formulate an algorithm for TSM. These algorithms were validated through inter-sensor comparisons and extensive sea-truth collection campaigns.

The objective of EOS-6 Ocean Colour Monitor (OCM) is to continue the legacy of Oceansat-1 & 2 OCM by providing the quantitative information on bio-optical constituents such as chlorophyll-a concentration, vertical diffuse attenuation coefficient of light ($K_d$) at 490nm, total suspended matter, etc., to the user community.

## *1.2 Global Scenario*

Satellite based ocean-colour measurement has emerged as one of the important area for gathering information on bio-geo-chemical variability of oceans on a global and regional scale. The launch of Nimbus-7 satellite in 1978 carrying Coastal Zone Colour Scanner (CZCS) heralded a new era in field of biological oceanography. The demise of CZCS in 1986 left a decade-long void in the continuity of ocean colour data from space. Subsequently, in mid-1990's three satellites were launched. MOS (Modular Optoelectronic Scanner), by the German Aerospace Agency (DLR) on-board IRS-P3 satellite, Ocean Colour and Temperature Sensor (OCTS) on-board Advanced Earth Observation Satellite (ADEOS) by Japan and SeaWiFS (Sea Viewing Wide Field-of-View Sensor) on-board OrbView2 by NASA. The SeaWiFS local area coverage (LAC) at 1 km spatial resolution and global area coverage (GAC) at 9 km spatial resolution data has been extensively used worldwide for numerous applications of ocean biology.

Currently, data from a number of ocean colour sensors launched by Japan, USA, Korea, China and Europe are available to the user community for global and regional studies. Foremost among these are two MODIS (Moderate Resolution Imaging Spectroradiometer) sensors launched by NASA on the TERRA and AQUA satellites, VIIRS (Visible Infrared Imager Radiometer Suite) on SUOMI–NPP and JPSS on MERIS (Medium Resolution Imaging Spectrometer), OLCI (Ocean Land Color Instrument) on Sentinel 3A & 3B by ESA.  Korea has launched two Geostationary Ocean Colour Imager (GOCI -I and GOCI-II). China currently has four ocean colour sensors providing data: Chinese Ocean Colour and Temperature Scanner (COCTS) with a spatial resolution of 1.1 km, Coastal Zone Imager (CZI) with a spatial resolution of 250 m and two Medium Resolution Spectral Imager (MERSI) onboard FY-3A and FY-3B. Japan launched the Second-Generation Global Imager (SGLI) on the Global Change Observation Mission-C (GCOM-C) in 2017.

Atmospheric correction techniques are based on radiative transfer equations, which describe the interactions between various atmospheric constituents and the sea surface that affect satellite measurements. For ocean colour sensors with bands limited to the NIR region, standard atmospheric correction procedure involving radiative transfer equations are applied to retrieve remote sensing reflectances. The classical 'black pixel' approach assumes negligible water leaving radiance in NIR bands and independently determines the aerosol radiance.  The SeaWiFS algorithm of Gordon and Wang (1994) computes the radiative transfer in the ocean-atmosphere system for a reference set of aerosol models. Multiple scattering is fully accounted in this algorithm, but the algorithm relies strongly on the choice of aerosol models. Fukushima et al. (2000) also developed an atmospheric correction algorithm for analysis of OCTS data using multiple scattering and aerosol model based approach, almost similar to the one used for SeaWiFS processing. However, standard atmospheric correction algorithms often fail over turbid coastal waters due to high backscattering of suspended sediment and other inorganic particulates (Mao et al. 2013).  this leads to non-zero water-leaving radiance in NIR, resulting in overestimation of the aerosol optical thickness in blue-green region of the electromagnetic spectrum, causing 'over-correction' in the visible part of spectrum. As a result, negative water leaving radiances at shorter wavelengths are generated in coastal waters. Ruddick *et al,* (2000) developed a method which assumes spatial homogeneity of 765:865 nm ratios of aerosol as well as water-leaving radiance whereby a user must interactively determine an aerosol type. This method solves for aerosol radiance and water leaving radiance simultaneously in the NIR and was successfully applied to highly scattering waters of the North Sea.

Bio-optical algorithms quantitatively relate variations in the concentration of bio-optical properties to ratios of radiance measurements at various wavelengths. The currently operational global ocean chlorophyll-a algorithm (Ocean Chlorophyll-4) is a cubic polynomial function based on band switching algorithm. It uses maximum remote-sensing reflectance ratio of 443, 490 and 510 nm to 555 nm band and has been developed based on an extensively calibrated and enhanced set of *in-situ* measurements.

## 2. Algorithms

The procedure for retrieving geophysical products relies on cloud free and glint free water pixels during daytime conditions over open ocean and coastal waters. Therefore, geophysical retrievals are performed only after applying land/cloud and sun-glint mask from OCM-3 TOA radiance. After atmospheric correction, remote sensing reflectance or normalized water-leaving radiance over water is retrieved. These are subsequently used in bio-optical algorithms for the retrieval of chlorophyll-a (Chl-a), and vertical diffuse attenuation coefficients ($K_d$).

### *2.1 Sun glint Characterization and masking*

Sun-glint is a phenomenon observed in ocean colour images due to the relative orientation of the sun and the sensor., The ocean surface, roughened by winds, acts as a strong specular reflector, reflecting the sun rays back to the sensor. The incoming radiation is specularly reflected back to the sensor without interacting with water constituents (Mohan *et al.* 2001). OCM-1 and OCM-2 used a payload steering mechanism to tilt the sensor by $\pm$ 20º to reduce glint contamination in images over the Indian subcontinent. The same strategy is also followed in OCM-3. However, due to its fixed orbit geometry, the global coverage of OCM-2 had permanent glint covered areas which shifted according to seasons causing data gaps in monthly binned images. This configuration of OCM-2 made it impossible to study basin scale properties for climate related applications. Therefore, in OCM-3, it was proposed to have a marching orbit to obtain glint free data every 5 to 6 days.

## 2.2 Atmospheric correction

In ocean colour remote sensing, radiance at the sensor level can be modeled using a molecular and particle scattering based approach and gaseous absorption in the atmosphere for a given solar and satellite geometry. Path radiance due to Rayleigh scattering computed for single scattering approximation is expressed in Gordon *et al.* 1988. The total signal received by sensor over ocean surface is the linear combination of atmospheric radiance (through scattering and absorption processes) and transmitted water leaving radiance from surface to sensor level. Basis of the algorithm is to remove the atmospheric effects from satellite data over water bodies to derive normalized water-leaving radiance/ remote sensing reflectance in the visible range. Atmospheric path radiance arises due to scattering of solar radiance by atmospheric gases and aerosols. Scattering of electromagnetic radiation by atmospheric gaseous molecules (Rayleigh scattering) occurs when light is elastically scattered by particles which are much smaller than the wavelength of incident light. Radiance resulting from Rayleigh scattering can be calculated using existing mathematical models for a given geometry using atmospheric pressure. Aerosols on other hand, are particles (liquid or solid) suspended in atmosphere which are much larger than the gas molecules. Optical properties of aerosol depend mainly upon their particle size distribution and refractive index.

Gordon and Wang in 1994 used the Shettle and Fenn, 1979 aerosol model to retrieve the aerosol optical thickness using two NIR bands. In the case of EOS-06 OCM, there will be two possibilities, first one, 780 and 870nm bands will be used for correcting atmosphere over open ocean while second one, 870 and 1010 nm bands will be used for correcting atmosphere over optically complex waters/ turbid waters.

$$L_t(\lambda) = L_r(\lambda) + L_a(\lambda) + L_{ra}(\lambda) + T(\lambda)L_g(\lambda) + t(\lambda)L_{wc}(\lambda) + t(\lambda)L_w(\lambda)$$

Where $L_r(\lambda)$, $L_a(\lambda)$, and $L_{ra}(\lambda)$ are the radiance contributions from multiple scattering of air molecules (Rayleigh scattering with no aerosols), aerosols (no molecules), and Rayleigh-aerosol interactions, respectively. $L_g(\lambda)$ is the specular reflection from the direct Sun (sun glint) radiance, $L_{wc}(\lambda)$ is the radiance at the sea surface resulting from sunlight and sky-light reflecting off whitecaps on the surface, and $L_w(\lambda)$ is the water leaving radiance. $T(\lambda)$ and $t(\lambda)$ are the atmospheric-direct and diffuse transmittances in the sensor-viewing direction respectively. The sun glint radiance can be written as:

$$L_g(\lambda) = F_0(\lambda)T_0(\lambda)L_{GN}$$

Where $F_0(\lambda)$ and $T_0(\lambda)$ are the extraterrestrial solar irradiance (adjusted for the Earth-Sun distance variations) and the atmospheric direct transmittance at the solar direction, respectively. $L_{GN}$ is the normalized sun glint radiance. For a given ocean color image pixel, the value of $L_{GN}$ depends on the solar and the viewing geometry, the sea-surface wind speed and the wind direction.

## *2.3 Aerosol models*

Gordon and Wang in 1994 used the aerosol models (Shettle and Fenn 1979) based on different particle size and relative humidity, which resulted in the corresponding change in their refractive indices. The particle size distribution is explained by bimodal log-normal distribution. Since particle size increases with increasing relative humidity (RH) from 0 to 99%. Three type of interpretations can be made based on the particle distribution, one is aiken or nucleation mode, second one is accumulation mode and third one is coarser mode. Further, under a broader category of aerosols,-two types of aerosols commonly occur in nature. Tropospheric aerosols based on its physical characteristics, fine-mode in nature and oceanic aerosols larger in nature. These two types of aerosols in variable proportion, take the form of some other aerosol categories namely maritime and coastal. Again, tropospheric aerosol consolidates the particle size fraction in the ratio of 7:3 i.e. (70% water soluble and 30% dust like particles) without oceanic contribution. This means the mode radii of tropospheric aerosols is always less than 0.1 um with single scattering albedo varies from 0.959 to 0.989 and for RH ranging from 0 to 98%. While for the larger fraction as a sea salt component in the oceanic aerosol type, the mode radii is greater than 0.3um as relative humidity (RH) increases from 0 to 98% with single scattering albedo considered to be unity [Gordon, 1997]. Therefore, $\epsilon(\lambda_l, \lambda_u)$ ceases the spectral variation due to swelling in particle size with increasing relative humidity as shown in Figure-1. The next category with the 99% particles of tropospheric nature and 1% oceanic pertains to the category of Maritime aerosol. Similarly, the coastal aerosol model comprises 99.5% tropospheric and 0.5% oceanic.

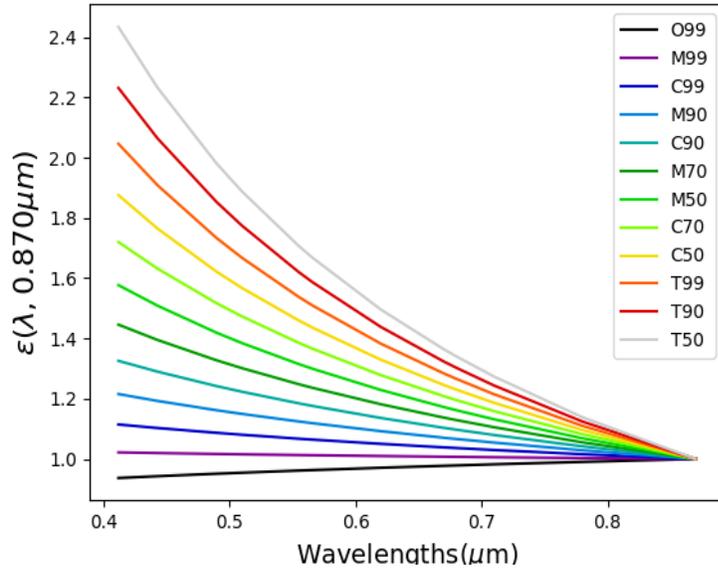

Figure-1 The look up table based on twelve aerosol model, represents the spectral variation in $\epsilon(\lambda_l, \lambda_u)$.

This model gets further utilized for retrieval of spectral remote sensing reflectance with aerosol optical thickness at 870nm.

## 3. Bio-optical algorithms

### *3.1 Chlorophyll-a concentration algorithm*

The development of ocean-colour Chl-*a* algorithm requires a large number of simultaneous *in-situ* radiance/reflectance and Chl-*a* data, preferably coinciding with satellite overpass (Gordon *et al*., 1983; Balch *et al*., 1992). Using the *in-situ* data collected from regional and global oceans, chlorophyll-a algorithms were developed for SeaWiFS , employing semi-analytical, semi-empirical and empirical approaches (O'Reilly *et al*., 1998, 2000). A large number of semi-analytical and empirical bio-optical algorithms were assessed using the *in-situ* data collected during ship campaigns in the Arabian Sea (Chauhan *et al.,* 2002). Among these, the OC2 algorithm performed well in estimating Chl-*a*. Thus, OC2 algorithm was subsequently adopted for routine processing of the ocean colour satellite images of Oceansat-1 OCM data. Further validation of Chl-*a* concentrations from OCM images and *in-situ* measured Chl-*a* resulted in good agreement, with an $R^2$ of 0.90 and RMS error of 0.125 mg m$^{-3}$. The total validation points were 43 and the range of Chl-*a* was from 0.072 to 5.9 mg m$^{-3.}$ The OC2 algorithm was developed using the *in-situ* database known as SeaBAM (N=919) (O'Reilly *et al*., 1998).

The NASA bio-optical Marine Algorithm data (NOMAD) is the largest global, high quality *in-situ* bio-optical dataset that is publicly available to develop and validate the satellite ocean-colour algorithms (Werdell and Bailey, 2005). It includes a wide range of Chl-*a* from 0.1 to 81.86 mg m$^{-3}$, covering a variety of mesotrophic and eutrophic waters. Chl-*a* was acquired using both fluorometric and HPLC methods (Werdell *et al*., 2003), with nearly two-thirds of the Chl-*a* measured using HPLC gave better estimations as compared to fluorometric method. Thus, all the fluorometrically measured Chl-*a* values were converted into HPLC Chl-*a* using the following relationship;

$$\text{HPLC (Chl)} = 0.8105 * \text{Fluor (Chl)} + 0.1931 \; (R^2 = 0.9)$$

NOMAD data comprises of more than 800 bio-optical measurements along with Chl-*a* ranging between 0.1 to 81.86 mg m$^{-3}$ (Werdell and Bailey, 2005). The remote sensing reflectance ratios for $R_{rs}443/R_{rs}555$, $R_{rs}490/R_{rs}555$ and $R_{rs}510/R_{rs}555$ were plotted against Chl-*a* for all the data points.

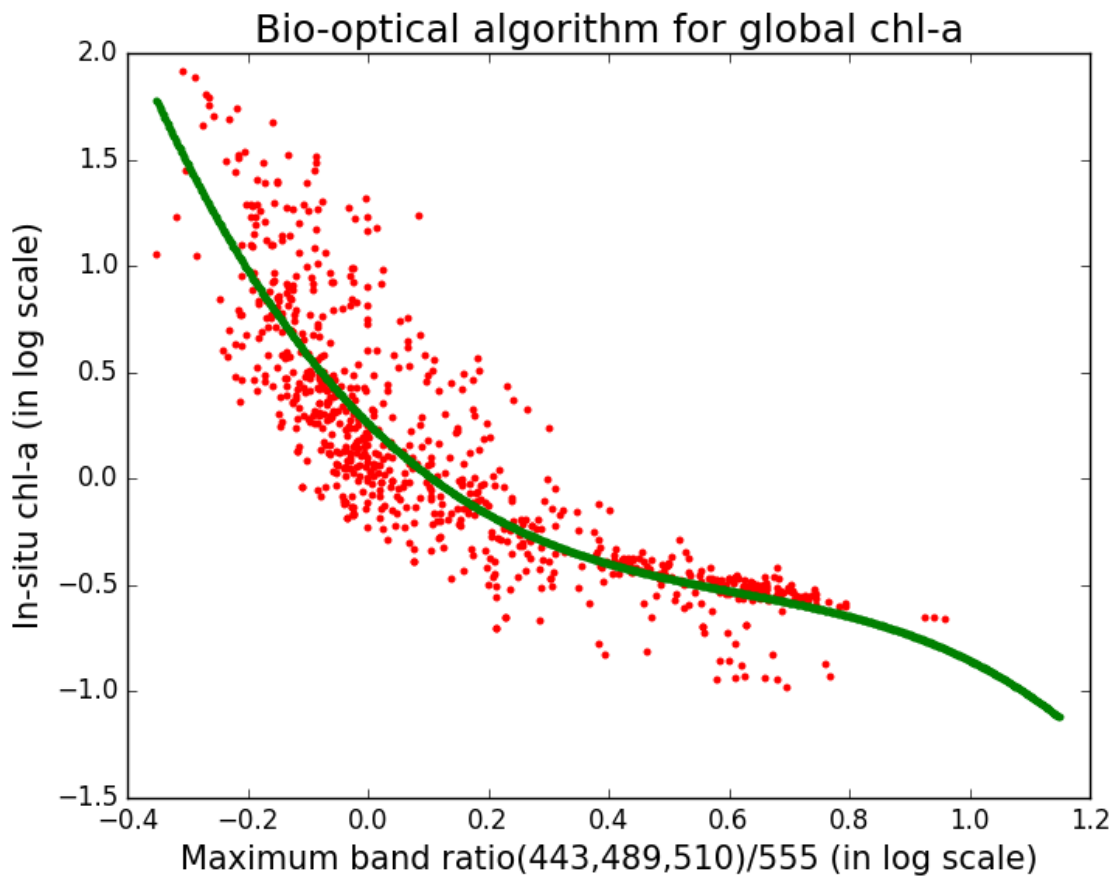

Figure-2 Graph shows the In-situ Chl-a in log-scale vs maximum band ratio in log-scale. Bio-optical algorithm for global chl-a was derived from NOMAD dataset

A robust relationship was obtained between maximum band ratio of $R_{rs}443/R_{rs}555$, $R_{rs}490/R_{rs}555$ and $R_{rs}510/R_{rs}555$ and corresponding Chl-*a*. The maximum $R_{rs}$ band ratios (MBR) verses Chl-*a* demonstrated a sigmoidal relationship as observed by many researchers (Morel and Maritorena, 2001). A cubic polynomial functional form was found to best fit the maximum band ratios (MBR) and Chl-*a* (Chauhan et al., 2002, 2003). The cubic polynomial formula yielded good results when tuned with log-transformed merged data set, with an $R^2 = 0.8$ and RMSE ~ 0.2. Cubic polynomial equation between log transformed Chl-*a* and MBR is given as

$$Log_{10}(C) = a+b*R+c*R^2+d*R^3$$

Where, C represents the Chl-*a*, R represents the logarithm of MBR of $R_{rs}443/R_{rs}555$, $R_{rs}490/R_{rs}555$ and $R_{rs}510/R_{rs}555$ and a, b, c, d and e are the regression coefficients as listen in Table 2. The developed cubic polynomial equation captures the inherent relationship between the *in-situ* reflectance band ratios and the Chl-*a*. Further refinement of the coefficients will be undertaken as more data is populated.

*Table 2: Bio-optical algorithm for Chl-a estimation, developed using NOMAD dataset.*

| Model | Input Band Ratio | Coefficients |
|---|---|---|
| $log_{10}(C)=a+b*R+c*R^2+d*R^3$ | $R = log_{10}(max[Rrs443, Rrs490, Rrs510]/Rrs555)$ | a=0.2604; b= -2.8025; c= 3.6626; d= -1.976 |

## 3.2 Vertical diffuse attenuation coefficient ($K_d$)

The vertical diffuse attenuation coefficient of oceanic water, an important parameter for understanding light availability, can be derived from ocean colour satellite data. Oceanic and coastal production of phytoplankton depends largely on light and the diffuse attenuation coefficient ($K_d$) plays a significant role in defining the presence of light versus depth. Ocean waters have been optically classified based on the relation between diffuse attenuation coefficient and the content of plant pigment. This knowledge is crucial in understanding the upper ocean warming, that occurs through absorption of solar irradiance (400-800 nm).

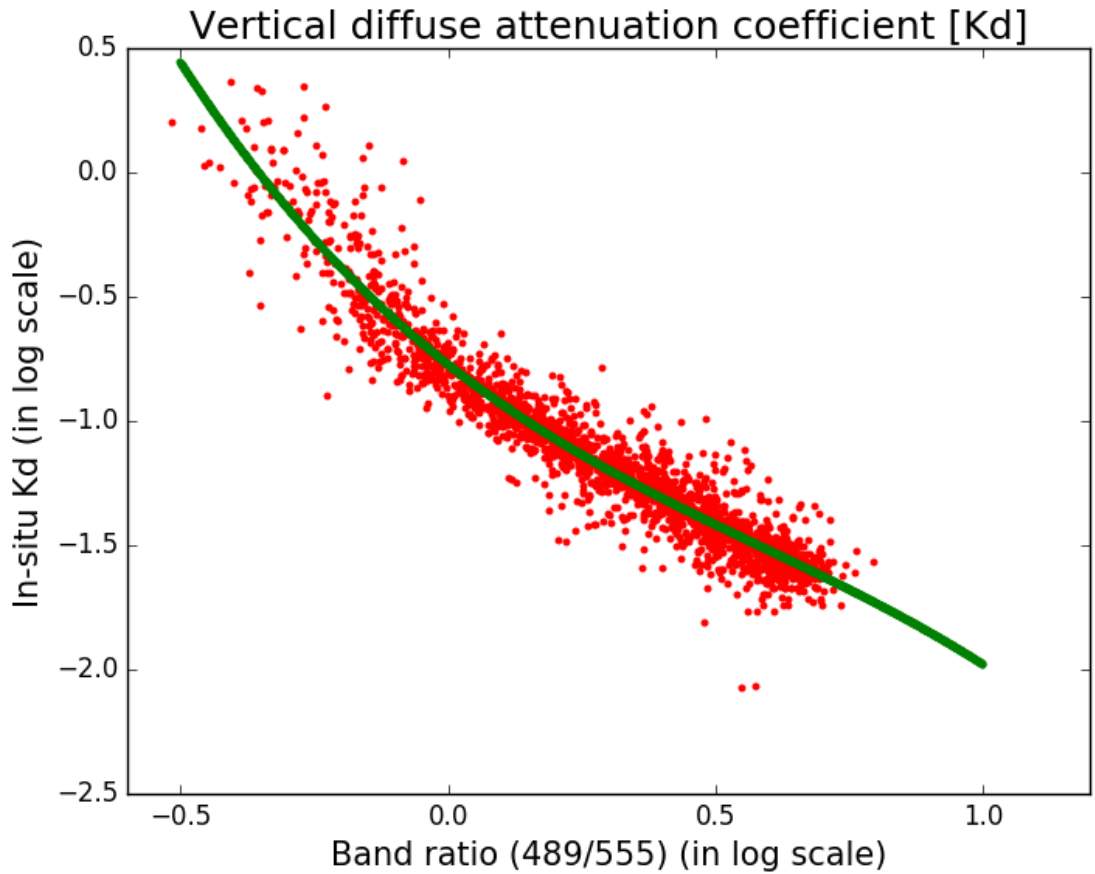

Figure-3 Graph shows the In-situ Kd in log-scale versus band ratio in log scale based on NOMAD dataset.

An algorithm has been developed to estimate $K_d$ (490 nm) as a function of ratio of water leaving radiance in 490 nm and 555 nm using the NOMAD merged data set. A cubic polynomial fit was found to best explain the functional relationship between the $K_d$ (490 nm) and ratio of remote sensing reflectance i.e. $R_{rs}(490)/R_{rs}(555)$. Following equation was obtained for the estimation of $K_d$ (490 nm) using NOMAD datasets with root mean square error of 0.09 and $R^2$ of 0.92 for log transformed data.

*Table 3: Algorithm for vertical diffuse attenuation coefficient $K_d$, developed using NOMAD dataset.*

| Model | Input Band Ratio | Coefficients |
|---|---|---|
| $Log_{10}(K_d\ 490) = a + b*K + c*K^2 + d*K^3$ | $K = \log_{10}[R_{rs}(490)/R_{rs}(555)]$ | a= -0.7732; b= -1.6961; c=1.141; d= -0.6511 |

Further refinement of the coefficients will be carried out as more data is populated.

## 4. Results and Discussion

The primary objective of this study is to demonstrate the importance of atmospheric correction in the processing of EOS-06 OCM data to derive oceanic bio-optical constituents like chl-a and vertical diffuse attenuation coefficient $K_d$ at 490nm. The atmospheric correction procedure involves the removal of effects caused by molecular and particle scattering, as well as gaseous absorption in the atmosphere. The paper discusses the use of different aerosol models for the atmospheric correction of Oceansat-3 data over open ocean and optically complex waters. Another objective of the study was to develop and validate bio-optical algorithms with the help of error propagation theory (GUM, 2008) for the EOS-06 OCM focusing on Chlorophyll-a concentration and vertical diffuse attenuation coefficient (Kd). The Chlorophyll-a concentration algorithm was developed based on maximum band ratio of (443,489,510) and 555nm with *in-situ* Chl-a data. We made use of cubic polynomial to capture the relationship between in-situ Rrs band ratios and Chl-a, which showed promising results, with an $R^2$ of 0.8 and an RMSE of approximately 0.2 for high dynamic range of chl-a i.e. 0.1 to 81 mg/m$^3$. Chl-a algorithm not only focusses on open ocean but also coastal region as well. To estimate diffuse attenuation coefficient, a cubic polynomial fit was used to relate Kd with the ratio of remote sensing reflectance(Rrs) at 490 nm and 555 nm. NASA bio-optical marine algorithm dataset (NOMAD) dataset (Werdell and Bailey, 2005) was used for this purpose to obtain a strong correlation ($R^2$ of 0.92) and low RMS error (~0.09) for log-transformed data.

In summary, the operational and bio-optical algorithms developed for the retrieval of geophysical products from EOS-06 OCM data have shown encouraging results. The algorithms have proven to be effective in accurately estimating Chl-a concentration and $K_d$, providing valuable insights into the ocean's bio-optical properties. These findings highlight the potential of EOS-06 OCM data for ocean colour studies and geophysical product retrieval. As we continue to accumulate more data, we plan to refine the algorithms further to improve the accuracy of the retrievals. OCM-3 (L1C) data on Feb 16, 2023 for path 53 and row 13 top of the radiance (L1C-LAC) data was used to retrieve remote sensing reflectance with aerosol optical thickness at 870nm as shown in Figure-4. Chl-a (mg/m$^3$) and vertical diffuse attenuation coefficient (1/m) at 490nm were derived using remote sensing reflectance based on Table-2 and Table-3 respectively as shown in Figure-4.

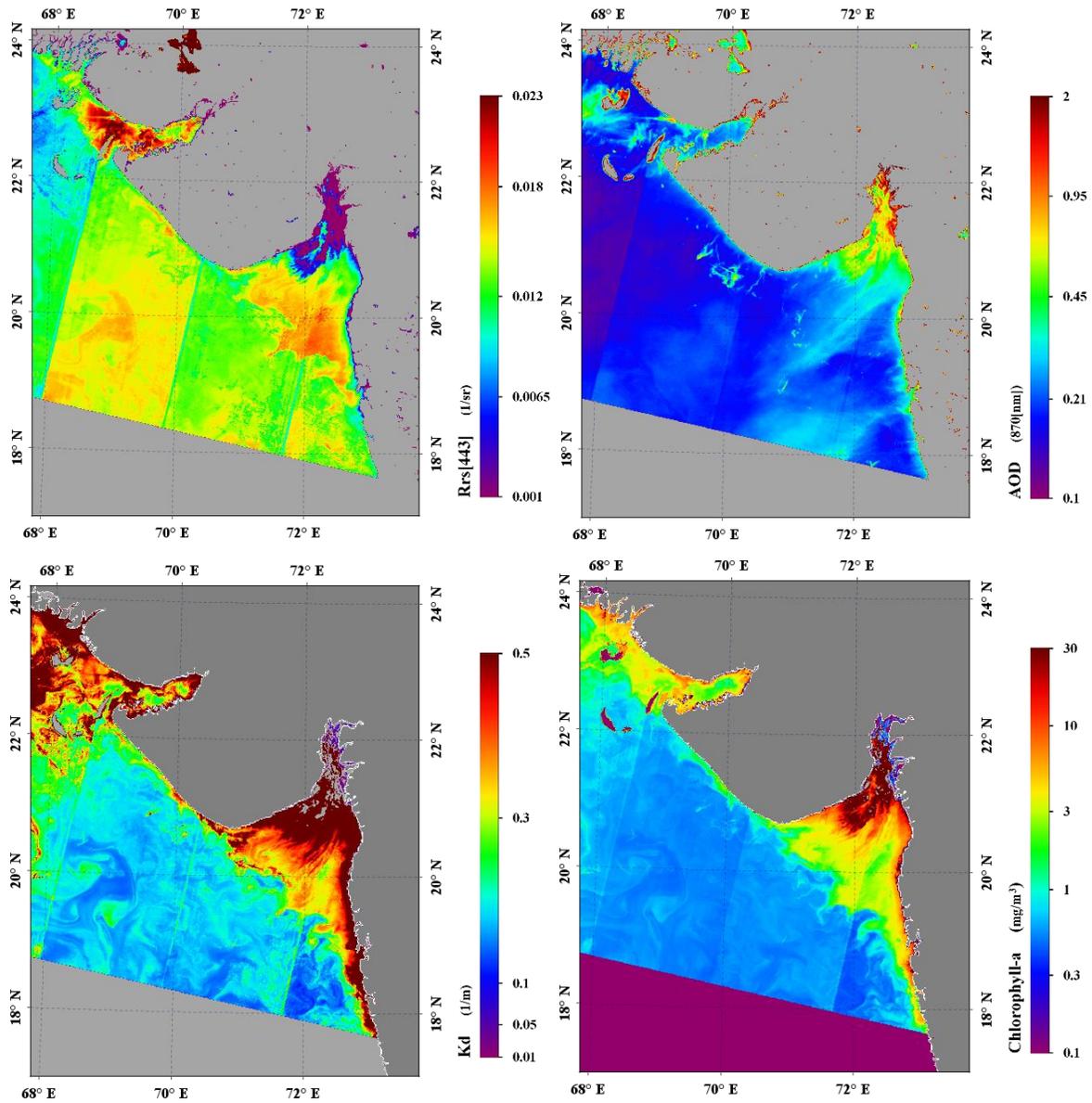

Figure-4 In top row, Remote sensing reflectance (1/sr) at 443nm, aerosol optical depth at 870nm and in bottom row, vertical diffuse attenuation coefficient (1/m) at 490nm and chlorophyll-a(mg/m$^3$) derived using EOS-06 (OCM).

## 5. Error propagation theory (GUM 2008)

Consider a model output y (y = f(x)) with uncertainty of U(y) and U(x) the uncertainty of any x, then according to the GUM 2008, the expression is:

$$U^2(y) = \sum_{i=1}^{n}\sum_{j=1}^{n} \frac{\delta f}{\delta x_i}\frac{\delta f}{\delta x_j} U(x_i, x_j) = \sum_{i=1}^{n}\left[\frac{\delta f}{\delta x_i}\right]^2 U^2 x_i + 2\sum_{i=1}^{n-1}\sum_{j=i+1}^{n}\frac{\delta f}{\delta x_i}\frac{\delta f}{\delta x_j} U(x_i, x_j)$$

The uncertainty associated with $x_i$ and $x_j$, can be expressed as

$$U(x_i, x_j) = R(x_i, x_j)U(x_i)U(x_j)$$

In the above expression, $R(x_i, x_j)$ represents the correlation coefficient between $x_i$ and $x_j$. If they are not correlated to each other then $R(x_i, x_j)$ becomes zero otherwise the second case persists as

$$U^2(y) = \sum_{i=1}^{n}\left[\frac{\delta f}{\delta x_i}\right]^2 U^2 x_i.$$

## 6. Uncertainty

Guide to the expression of uncertainty in measurement (GUM, 2008) focusses on error propagation theory. With its help, uncertainties associated with geophysical products as chlorophyll-a (mg/m³) and vertical diffuse attenuation coefficient(1/m) derived from ocean colour radiometry, can be estimated as shown in Figure-5. Since no measurement is complete without error, so it's obvious to interpret the error associated with product level on pixel to pixel basis. As per the International Ocean Colour Coordinating Group (IOCCG), relative uncertainty associated with top of the atmosphere(TOA) radiance is 0.5% in blue-green bands. Since water leaving radiance contribution to the top of the atmosphere is ~10% of TOA radiance. So 0.5% uncertainty at TOA gets translated to 5% uncertainty in remote sensing reflectance over open ocean as shown in Figure-5 first row. In the figure, error is more i.e. >5% in the open ocean because of dominance of bloom like biomass results strong absorption in blue channel. That means contribution from blue band (443nm) to sensor level is <10%. While in coastal region, error exceeds 5% in many occasions very easily. Colour dissolve organic matter (CDOM) dominated waters always possess retrieval accuracy problem because the contribution from CDOM dominated water is <1% to TOA level.

Similarly, following the international norm (IOCCG), 5% uncertainty in remote sensing reflectance was assimilated into the bio-optical model to examine the error distribution on pixel to pixel basis in the product i.e. Chl-a as shown in second row of Figure-5. It was concluded that uncertainty in derive Chl-a from EOS-06 OCM i.e. <35% in open ocean follows global norm as per the IOCCG. Uncertainty in vertical diffuse attenuation coefficient at 490nm exceeds >15% sometime in open ocean because of the same problem as discussed above.

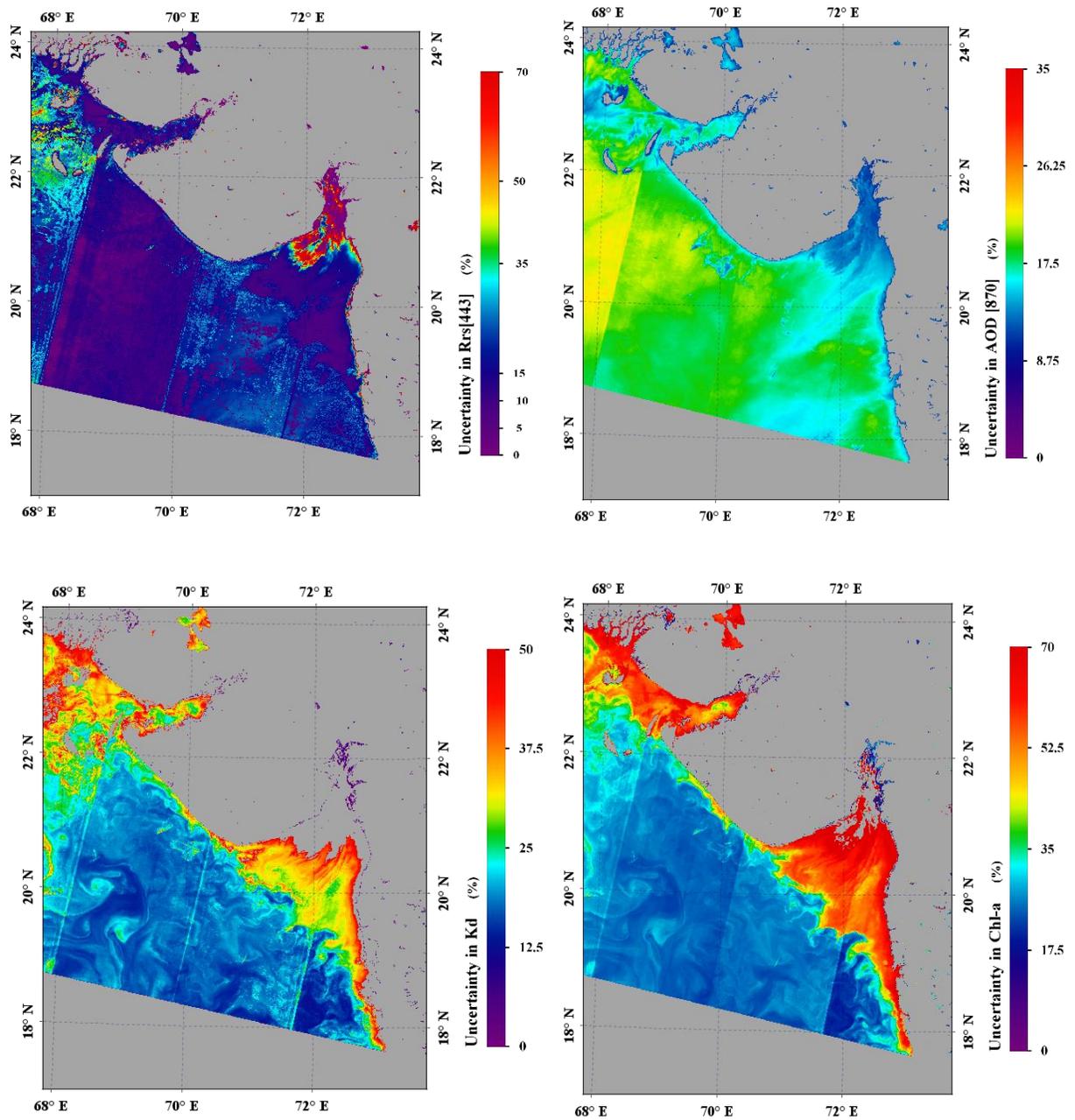

Figure-5 In top row, uncertainty in Remote sensing reflectance (1/sr) at 443nm, uncertainty in aerosol optical depth at 870nm and in bottom row, uncertainty in vertical diffuse attenuation coefficient (1/m) at 490nm and uncertainty in chlorophyll-a(mg/m$^3$) derived using EOS-06 (OCM).

## 7. Conclusions

In this paper, algorithms for bio-optical characterization i.e. chlorophyll-a (mg/m$^3$) and vertical diffuse attenuation coefficient i.e. $K_d$ (1/m) were developed using NASA bio-optical marine algorithm datasets (NOMAD) with correlation coefficient $R^2$ (>0.8) and (>0.9) for global waters respectively. They were successfully derived from remote sensing reflectance using EOS-06 OCM. Since remote sensing reflectance was retrieved from EOS-06 OCM over open ocean in Arabian sea with uncertainty of 5 to <10%. While it exceeds 10% in the coastal and high bloom conditions in open ocean also. The uncertainties in ocean colour radiometry propagate in the products i.e. chl-a and Kd with <35% and <20% in open ocean respectively.

## Acknowledgements

We acknowledge Director, Space Applications Centre (ISRO), Ahmedabad for his continuous support for pursuing this study. We also acknowledge Director, National remote Sensing Centre (NRSC) for providing data for the study. We are thankful to Dr. Rashmi Sharma, Deputy Director (EPSA) and Project Director (Application (EOS-06)) SAC for her kind support and motivation.We are also thankful to Dr. Pradeep Thapliyal, Associate project Director (Application (EOS-06)) and Head (ASD-SAC) and Dr. Bimal Bhattacharya, Group Director (AESG-SAC) for their valuable suggestions and guidance during the work. We are also thankful to Deputy Director (SEDA-SAC) and Deputy Director (SIPA-SAC) for their useful suggestions during OCM-3 review. Thanks are also due to all those who supported directly or indirectly for this work.

**Data Availability Statement**  The ISRO's data EOS-6 (OCM) and NASA bio-optical marine algorithm datasets (NOMAD) have been used in the manuscript, that can be accessed from the following: https://seabass.gsfc.nasa.gov/wiki/NOMAD/nomad seabass v2. a 2008200.txt

## Author statement

Anurag Gupta contributed in manuscript for conceptualization, methodology, software and writing the manuscript. Mr. Debojyoti Ganguly contributed in writing the manuscript. Dr. Mini Raman contributed in writing review and editing of the manuscript with her supervision. Dr. K.N. Babu